\documentclass[aps,twocolumn,prl,10pt]{revtex4-1}
\usepackage{silence}
\usepackage{bm}
\usepackage{url}
\usepackage{amsmath,amssymb,amsfonts,graphicx,float}
\renewcommand{\vec}[1]{{\mathbf{#1}}}
\newcommand{\beq}{\begin{eqnarray}}
\newcommand{\eeq}{\end{eqnarray}}
\usepackage{amsmath}
\usepackage{amssymb}
\usepackage[paper=letterpaper,margin=1in]{geometry}
\usepackage{braket}
\usepackage{upgreek}

\renewcommand{\vec}[1]{\boldsymbol{#1}}

\def\phi{\varphi}

\usepackage{tikz}

\begin{document}

\title{Strongly Coupled Fixed Point in $\varphi^4$ Theory}
\author{ Anthony Hegg}
\author{Philip W. Phillips}
\thanks{Guggengeim Fellow}

\affiliation{Department of Physics and Institute for Condensed Matter Theory,
University of Illinois
1110 W. Green Street, Urbana, IL 61801, U.S.A.}

\date{\today}

\begin{abstract}
We show explicitly how a strongly coupled fixed point can be constructed in scalar  $g\varphi^4$ theory from the solutions to a non-linear eigenvalue problem.  The fixed point exists only for $d< 4$, is unstable and characterized by $\nu=2/d$ (correlation length exponent), $\eta=1/2-d/8$ (anomalous dimension). For $d=2$, these exponents reproduce to those of the Ising model which can be understood from the codimension of the critical point. At this fixed point, $\varphi^{2i}$ terms with $i>2$ are all irrelevant. The testable prediction of this fixed point is that the specific heat exponent vanishes. 2d critical Mott systems are well described by this new fixed point.
\end{abstract}

\maketitle

In 1976, Benzi, Martinelli and Parisi\cite{parisi} stated at the outset of their paper that ``It is well known that in most of the interesting cases of field theory the perturbation expansion in the coupling constant is useless.''  Since  then, little progress has enabled controlled computation in the strongly coupled regime where standard perturbative methods break down.  For example even in the simplest case of scalar $\varphi^4$ theory, it has proven notoriously difficult\cite{luscher, bender,frasca,kogut,wilsonanom} to establish a fixed point, the terminus of a renormalization group flow, at strong coupling.  

This state of affairs is unfortunate because numerous physical systems abound in which the interactions dominate, the normal state of the copper-oxide superconductors and bound states of quarks, two cases in point.  In the context of the former problem, the most widely used technique is the dynamical mean-field theory (DMFT)\cite{dmft} in which the physics of an extended system is boot-strapped from the ultra-local physics of a single site or a cluster.  While this procedure gives experimentally accurate results even at low energy, no fundamental principle, such as the variational principle, underlies DMFT.  What is peculiar is that even some of the results\cite{faulkner} from the application of the gauge-gravity duality to fermionic matter at finite density have obtained spectral functions that exhibit the ultra-local scaling of DMFT.  This coincidence would be more than an accident if perhaps strongly correlated systems, such as Mott insulators, are controlled by an ultra-local fixed point and hence any numerical scheme that builds in local physics must flow to the strongly coupled fixed point. 

In this paper, we demonstrate within scalar $\varphi^4$ theory that such an ultra-local fixed point actually exists and we show how it can be accessed by expanding in the exact eigenstates of a non-linear eigenvalue problem.   In the standard perturbative approach to scalar $\varphi^4$ theory,  the scaling dimension of $[\varphi]=(d-2)/2$ is chosen so that the kinetic term has zero scaling dimension. Since perturbation theory in the coupling constant  fails when the interactions dominate, we seek an alternative formulation. The motivation for our approach begins by simply noting that if the engineering scaling dimension of the $g\varphi^4$ term is set to zero by choosing $[\varphi]=d/4$, then the kinetic energy term now has scaling dimension $2+d/2$ and hence would be relevant only for $d>4$! Consequently, to determine the physics for $d<4$, it would then makes sense to treat the kinetic term as a perturbation and a fixed point must exist as $g\rightarrow \infty$. Nonetheless, this conclusion is tenuous  because we don't expect scaling at such strongly coupled fixed points to correspond to simple engineering dimensions. Our goal is now clear: find an unequivocal fixed point at strong coupling and subsequently calculate scaling exponents in its vicinity, but we must do so without considering any term to be a perturbation from the start.

To solve this problem, we expand the action 
\beq
\label{action}
S = -\int d^d x \left\{-\phi \nabla^2 \phi + r \phi^2 + g \phi^4 \right\}
\eeq
in terms of the solutions to the non-linear eigenvalue equation 
\beq
\label{phiEqn}
-\nabla^2 \phi + r \phi + g \phi^3 = \lambda \phi
\eeq
with $\lambda$ the associated eigenvalue.  Note this is not the equation of motion, but simply an eigenvalue equation that provides a complete set of states for the expansion of the action.  We impose periodic boundary conditions $\phi(0) = \phi(L)$ and $\phi'(0) =\phi'(L)$.
 The exact solution to this equation
\beq
\phi_n &=& c_n \text{sn}(p_n \cdot x + \theta, m_n) \\
\label{eqn:eigensol1}
\lambda_n &=& p_n^2 + r + \frac{gc_n^2}{2} \\
\label{eqn:eigensol2}
m_n &=& \frac{gc_n^2}{2 p_n^2}\\
\label{eqn:eigensol3}
p_n &=& \frac{4 K(m_n)n}{L}
\label{eqn:eigensol4}
\eeq
forms a complete\cite{craven} nonlinear basis in terms of the Jacobi elliptic functions, $\rm sn(z,m)$.  Here $n\equiv \vec{n}$ is a vector of integers and we assume each dimension has length $L$ such that $p_n^2 = \sum_{j=1}^d p_{n_{x_j}}^2$, and $\theta=0$ or $\theta=2K(m_n)$ gives the odd or even solutions respectively with $K(m)$ being the complete elliptic integral of the first kind. 

The utility of the elliptic functions is that they encode the interactions non-perturbatively. In general, however, an expansion of the action in this basis yields complicated cross-terms between the nonlinear eigenfunctions. As will be self-consistently shown below, there are two cases corresponding to fixed points at $(r,g)=(0,0)$ and $(r,g)=(0,\infty)$ where the diagonal terms in the action dominate over the cross-terms resulting in a simple expression of the action near these points
\beq
\label{diagAction}
S_n = -\int_0^L \lambda_n \phi_n^2
\eeq
where $n_{x_j}=n \forall j$ which simplifies $
p_n^2=d~p_{n_{x_1}}\equiv d~p^2$,
where we have dropped the index notation. 

Near these two points in coupling constant space the cutoff portion of the action can be shown to be positive definite, see the supplemental material for details. Factoring out of the integrand the large quantities $p_{n_0}^2$ or $g$ respectively one can perform the method of steepest descent to integrate out the cutoff amplitude about the global minimum $c_{n_0}=0$. Since the global minimum vanishes near these two points, the terms in the action with the most copies of the amplitude dominate the behavior of the system. This along with analytic tractability motivates our choice in Eq.(\ref{diagAction}) and integration of the cutoff term is decoupled from the remaining degrees of freedom.

The result must be rescaled, which is initiated by having the momentum satisfy $k'=bk$ where $b>1$ and $k=2\pi n/L$. Assuming a general form for $p$ scaling, namely,
\beq\label{pscale}
p' = b^{d_p}p,
\eeq
we look at the gradient term in the action
\beq
\int d^d x p^2 c^2 \text{sn}(px,m)
\eeq
and find that, since the $\text{sn}$ part cannot scale, the eigenfunction amplitude scales as
\beq\label{cscale}
c' = b^{\frac{d-2d_p}{2}}c.
\eeq
We treat this amplitude scaling as analogous to the field scaling from traditional perturbative methods. The constraint on $m$ given in Eq.(\ref{eqn:eigensol3}) must scale as well, and since the only parameter free to scale in this equation is $g$ we use $m$-scaling to determine how $g$ scales. We first redefine our coupling constant as 
\beq
m &=& \frac{g c^2}{2dp^2} \\
\label{mDef}
m' &=& \frac{g' c^2}{2dp^2}.
\eeq
Using Eqs.(\ref{pscale},\ref{cscale}) we find that
\beq
\frac{g'c'^2}{2dp'^2} &=& \frac{g' c^2}{2dp^2}b^{d-4d_p} \\
&=& m' b^{d-4d_p}.
\eeq
A further constraint on $m$ is given by the periodic boundary condition Eq.(\ref{eqn:eigensol3}) along with the definition $k=\frac{2\pi n}{L}$ resulting in
\beq
p &=& 4K(m)\frac{k}{2\pi} \\
p&=&\frac{2K(m)}{\pi}k \\
p'&=&\frac{2K(m'b^{d-4d_p})}{\pi}bk \\
\Rightarrow p'&=&\frac{K(m'b^{d-4d_p})}{K(m)}bp\equiv b^{d_p}p.
\eeq
Rearranging the last equality we arrive at our rescaling for $m$
\beq\label{mscale}
m'=b^{4d_p-d}K^{-1}(b^{d_p-1}K(m)).
\eeq
Absorbing the remaining rescaled terms from the eigenvalue $\lambda$ into the rescaling of $r$, we find our final rescaling equation
\beq\label{rscale}
r'=b^{2d_p}r+b^{2d_p}\left( m-K^{-1}(b^{d_p-1}K(m)) \right).
\eeq

We can use Eqs. (\ref{mscale},\ref{rscale}) to identify fixed points (FP) of the theory. At such a point, the rescaling equations must simplify to $r' =r$ and $m'=m$. Two such points arise corresponding to $m \rightarrow 0$ and $m \rightarrow 1$, which we denote as the Gaussian (G) FP and the strongly coupled (SC) FP respectively. This rescaling method does not access the $d=3$ critical point of the theory, so we will not discuss this FP further. Applying these limits to the rescaling equations, we find
\beq\label{lims}
\lim_{m \to 0} m_{G}'&=&b^{d_p-1}m \\
\lim_{m \to 0} r_{G}'&=&b^{2d_p}r \\
\lim_{m \to 1} m_{SC}'&=&b^{4d_p-d}m \\
\lim_{m \to 1} r_{SC}'&=&b^{2d_p}r.
\eeq
For the Gaussian fixed point, it is straightforward to see that $g \rightarrow 0$ as $m \rightarrow 0$ from Eq.(\ref{eqn:eigensol3}) using the fact that $K(m=0)=\pi/2$ in Eq.(\ref{eqn:eigensol2}). To find the value of $g$ for the strongly coupled fixed point we first use Eq.(\ref{eqn:eigensol3}) to find that $\lim_{m \to 1} K(m) \rightarrow \infty$ implies that $\lim_{m \to 1} p \rightarrow \infty$. We then solve Eq.(\ref{eqn:eigensol2}) for g and assume the amplitude $c$ is finite to obtain $\lim_{m \to 1} g \rightarrow \infty$. In both limits we find that the $m$-dependence in Eq.(\ref{rscale}) vanishes to give the same rescaling equation shown in Eq.(\ref{lims}), which leads to $r^{*}=0$ for both fixed points. Therefore the fixed points $(r^{*},g^{*})$ we identify here correspond to the Gaussian $(0,0)$ and a new $(0,\infty)$ fixed point at strong $g \rightarrow \infty$ coupling. Using Eqs. (\ref{lims}) we immediately find the required values of $d_p$ for each fixed point as
\beq\label{dp}
d_{p,G} &=& 1 \\
d_{p,SC} &=& \frac{d}{4}.
\eeq

\begin{figure}[t]
\centering
\includegraphics[scale=0.5]{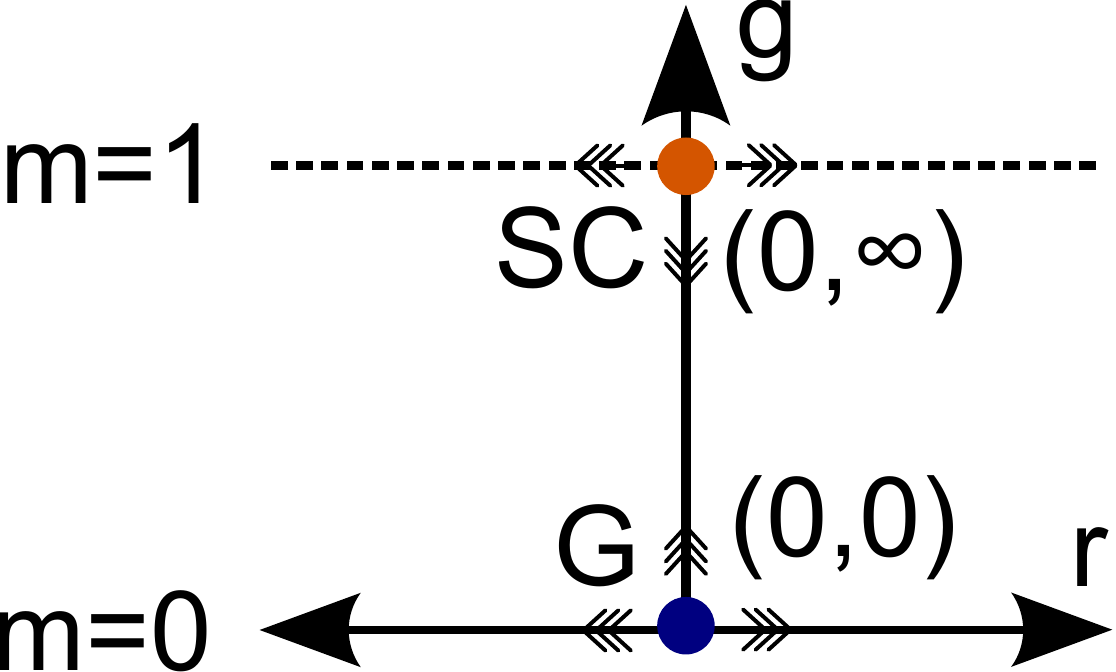}
\caption{Flow diagram in the vicinity of Gaussian $(0,0)$ and strongly coupled $(0,\infty)$ fixed points. Axes are given as $r$ vs $g$ with corresponding values of $m$ on the left. The dotted line represents the $m \rightarrow 1$ or $g \rightarrow \infty$ limit. Since flows in both cases are all away from each fixed point both points are unstable.}
\label{flow}
\end{figure}

Before calculating the power-law exponents for each of these points we characterize them based on the rescaling flows in their vicinity. We can do this by choosing $r^{*}+\delta r$ and $m^{*} + \delta m$ to be an infinitesimal shift away from the corresponding fixed point while using the values for $d_p$ obtained at that fixed point. We then apply Eqs. (\ref{mscale},\ref{rscale}) to find out in which direction the new values flow. Taking $m=0+\delta m$ at the Gaussian FP and $m=1-\delta m$ at the SC FP, we find
\beq
m_{G}' &=& b^{4-d} \delta m \\
m_{SC}' &=& K^{-1}\left( b^{\frac{d-4}{4}} K(1-\delta m) \right).
\eeq
As long as $d<4$ we find that for the Gaussian FP $m'>m$.  The opposite is true at the strongly coupled fixed point for $d<4$.  In this case,
$b^\frac{d-4}{4}<1$ and
$K(m)$ is a strictly increasing function resulting in $m'<m$.  At $d=4$, $m'=m$ and no non-trivial solution exists at strong coupling.  Our conclusion that a strongly fixed point exists therefore requires $d<4$.  This is consistent with mean-field behavior obtaining for $d\ge 4$. For either FP the equation for $r$ is simple and flows away from $r^{*}=0$ in both directions. These fixed points and their corresponding flows in $d=3$ are given in Fig. \ref{flow}.

In order to calculate the exponents for these fixed points, we first use the definition of $\nu$ as the inverse of the scaling for $r$. Recall from Eq. (\ref{dp}) that this is made especially simple given that all $m$-dependence drops out of Eq. (\ref{rscale}) for both fixed points. In both cases
\beq
\nu = \frac{1}{2d_p}
\eeq
and the problem reduces to identifying $d_p$ for each fixed point (see Eq. (\ref{dp})) resulting in 
\beq
\nu_{G} &=& \frac{1}{2} \\
\nu_{SC} &=& \frac{2}{d}.
\eeq

Typically at least two exponents are needed to fully quantify the exponents at a given fixed point with the rest determined using scaling laws\cite{kenna}. For the second exponent we use the definitions of $\eta$ as the difference between the field scaling at the given fixed point and that at the Gaussian fixed point. This gives $\eta=0$ by definition at the Gaussian point, but we can incorporate this result into a general formula as follows. In our eigensolution the field scales as the amplitude $c$ of the eigenfunctions. This scaling is fully determined by $d_p$ as shown in Eq.(\ref{cscale}). Setting the Gaussian value to $d_p=1$ we find that
\beq
\eta &=& \frac{d-2d_p}{2} - \frac{d-2}{2} \\
\Rightarrow \eta &=& \frac{1-d_p}{2}
\eeq
and for the strongly coupled fixed point, we obtain
\beq
\eta = \frac{1}{2} - \frac{d}{8}.
\eeq
Table \ref{tab:exponents} summarizes all the exponents.

\begin{table}[ht]
\caption{$\phi^4$  exponents for the Gaussian and strong coupled ($r=0$, $g\rightarrow \infty$) fixed points.  The exponents for the latter are valid strictly for $d<4$.} 
\centering 
\begin{tabular}{c c c} 
\hline\hline 
$\text{FP}$ & $\text{G}$ & $\text{SC}$ \\ [0.5ex] 
\hline 
$\nu$ & $\frac{1}{2}$ & $\frac{2}{d}$ \\ 
$\eta$ & $0$ & $\frac{1}{2}-\frac{d}{8}$ \\
$\alpha$ & $0$ & $0$ \\
$\beta$ & $\frac{1}{2}$ & $\frac{7}{8}-\frac{3}{2d}$ \\
$\gamma$ & $1$ & $\frac{3}{d}+\frac{1}{4}$ \\
$\delta$ & $3$ & $\frac{9d+12}{7d-12}$ \\ [1ex] 
\hline 
\end{tabular}
\label{tab:exponents} 
\end{table}

The SC fixed point in $d=3$ is an unstable one similar to the Gaussian FP as is evident from the flow diagram in Fig. (\ref{flow}). Since the exponents are obtained from the exact eigenstates and such states form a complete basis\cite{craven}, we have exactly characterized the strongly coupled fixed point.

A surprising consequence of the strongly coupled fixed point is that the exponents in $d=2$ reduce exactly to those of Onsager's in the 2d Ising model. This implies that the fixed point we have found here should be applicable quite generally to systems in which the interactions dominate. Of course for $d=2$ it is possible that operators other than $\varphi^4$ are relevant and hence a careful analysis of this system includes higher order terms. However, the codimension of the Ising critical point is 2\cite{goldenfeld}  (two relevant directions) and these are in general the quadratic strength $r$ and the external applied field $H$. If all of the remaining coupling parameters are found to be irrelevant then the universality class found here will remain unchanged. To check this we add terms such as $g_6 \phi^6$, $\cdots$ ,$g_{2i} \phi^{2i}$ that obey the Ising symmetry to the action. The nonlinear eigenvalue procedure used to generate the complete basis for the $g_4 \phi^4$ theory above generalizes to the $g_{2i} \phi^{2i}$ theory as well. We define this $\phi^{2i}$ hyperelliptic function by the inverse of the hyperelliptic integral

\beq
x = \int_{0}^{\phi} \frac{dt}{\sqrt{(1-t^{2})(1-m_{1} t^{2})\cdots(1-m_{i-1} t^{2})}}
\eeq
where
\beq
\vec{m} = (m_{1},m_{2},\ldots,m_{i-1})
\eeq
and $\phi=\text{sn}_{2i}\left(x,\vec{m}\right)$.
The general solution to the nonlinear eigenvalue problem is given by

\beq
\label{Sol2i}
\phi_{n}(\vec{x}) = c_{n}~\text{sn}_{2i}\left(\vec{p}_{n} \cdot \vec{x} + \theta_{n},\vec{m}_{n}\right)
\eeq
and the periodic boundary condition is satisfied by

\beq
\vec{p}_{n} = \frac{4 K_{2i}(\vec{m}_{n}) n}{L}
\eeq
where $K_{2i}(\vec{m}_{n})$ is the hyperelliptic generalization to the complete elliptic integral of the first kind

\beq
\int_{0}^{1} \frac{dt}{\sqrt{(1-t^{2})(1-m_{1n} t^{2})\cdots(1-m_{(i-1)n} t^{2})}}
\eeq
Inserting Eq.(\ref{Sol2i}) with $i=3$ into the nonlinear eigenvalue equation for $\phi^{6}$ and equating like terms we find that

\beq
\frac{g_4 c^2}{2 p^2} &=& \tilde{m} \\
-\frac{g_6 c^4}{3 p^2} &=& \bar{m} \\
\lambda &=& p^2 + r + p^2(m_1+m_2) \\
\Rightarrow \lambda &=& p^2 + r + \frac{g_4 c^2}{2} + \frac{g_6 c^4}{3}
\eeq
where $\tilde{m} = m_1 + m_2 + m_1 m_2$ and $\bar{m} = m_1 m_2$ and a similarly determined set of solutions obtain for the $\phi^{2i}$ case. Letting $m_1 \rightarrow 1$ and $m_2 \rightarrow 0$ while their product $m_1 m_2 \rightarrow 0$ leads to the desired FP location where $\bar{m} \rightarrow 0$ and $\tilde{m} \rightarrow 1$. We then generate the rescaling equation analogous to Eq.(\ref{mscale})

\beq\label{m6scale}
K_6(\tilde{m}'b^{d-4d_p},\bar{m}'b^{2d-6d_p})=b^{d_p-1}K_6(\tilde{m},\bar{m})
\eeq
and we see that $d_p=d/4$ and $d=2$ gives

\beq
K_6(\tilde{m}',\bar{m}'b)=b^{-1/2}K_6(\tilde{m},\bar{m})
\eeq
so that $K_6$ and $\bar{m}$ are both reduced upon rescaling showing that the $\phi^{6}$ term is irrelevant. In general coefficients with $i > 2$ are irrelevant, supporting the claim that this is indeed the $d=2$ Ising critical point.

The testable prediction of this strongly coupled fixed point is the value of the specific heat exponent. Because of the hyperscaling relation, $2-\alpha=d\nu$, our computed value for $\nu=2/d$ implies that $\alpha=0$ as shown in Table (\ref{tab:exponents}). Consequently, the divergence is at best logarithmic. Two independent systems seem to exhibit this behavior. First, in the pnictides, a logarithmic divergence of the form $\ln|x-x_c|$ of the specific heat in BaFe$_2$(As$_{1-x}$P$_x$)$_2$ has been seen in low fields\cite{carringtonhc,hcsiabrahams,matsudahc}. A direct measurement of $\alpha$ would be preferable rather than in inference based on the effective mass since the very meaning of a quasiparticle is obscured in the local limit. In addition, care must be taken to distinguish a pure $\ln|T|$ dependence from $T^a\ln|T|$ as is observed in many non-Fermi liquid systems\cite{millis,continentino,stewart} in which $\alpha\ne 0$. Second, a recent scaling theory of the finite temperature Mott transition\cite{mottcriticality} has predicted that the heat capacity only has a $\ln|T|$ dependence and as a result is well described by the $d=2$ Ising exponents. What our work clarifies is that $\alpha=0$ is a generic feature of a strongly coupled fixed point (for $d<4$) not just the $d=2$ Ising model.  The applicability to Mott criticality is expected as such systems are governed by strong local interactions.

\textbf{Acknowledgment:} P. Phillips thanks Sung-Sik Lee for an inspiring discussion at the Isaac Newton Institute from which the idea of scaling to the interactions was borne and S. Kachru and Brian Swingle for comments on earlier drafts of this manuscript. A. Hegg thanks B. Brinkman, W. Evans, and G. Vanacore for stimulating discussion leading to the rescaling process. A. Hegg and P. Phillips are supported by NSF DMR-1104909 which grew out of earlier work funded by the Center for Emergent Superconductivity, a DOE Energy Frontier Research Center, Grant No.~DE-AC0298CH1088.

\newpage
{\bf Supplemental Material: Decoupling at the Strongly Coupled Point}
Here we show that the partial trace over the cutoff degrees of freedom can be decoupled from the remaining degrees of freedom for $(r,g)\rightarrow (0,0)$ and $(r,g)\rightarrow (0,\infty)$. Expanding the action Eq.(\ref{action}) in the nonlinear basis Eq.(\ref{eqn:eigensol1}-\ref{eqn:eigensol3}), we focus our attention on the terms in the action containing a cutoff degree of freedom

\begin{widetext}
\beq
\text{S}_{n_{0}} = \int d^d x~ \phi_{n_{0}} \left\{ \left( -\nabla^2 + r \right)\sum_{n=1}^{n_{0}} \phi_n + g \sum_{n,p,q=1}^{n_{0}} \phi_n \phi_p \phi_q \right\}.
\eeq
Although any cutoff index $n_{0}$ can be chosen, for simplicity here we choose $n_{0}=2^m$ for $m=1,2,3,\ldots$. Each power-of-two basis function is orthogonal to all those functions of lower index, simplifying the action

\beq
\label{cutoffAction}
\text{S}_{n_{0}} &=& \int d^d x~ \phi_{n_{0}} \left\{ \left( -\nabla^2 + r \right) \phi_{n_{0}} + g \sum_{p,q=1}^{n_{0}} \phi_{n_{0}} \phi_p \phi_q \right\}.
\eeq
\end{widetext}
The quartic term is even more simple than it appears in Eq.(\ref{cutoffAction}). Due to orthogonality conditions, the only nonvanishing terms are

\beq
g \sum_{p,q}^{n_{0}} \phi_{n_{0}}^2 \phi_p \phi_q = g \left(  \phi_{n_{0}}^4 + \phi_{n_{0}}^2 \sum_{p,q=1}^{n_{0}-1} \phi_p \phi_q \right)
\eeq
where $\phi_p$ and $\phi_q$ are nonorthogonal. Although this basis is nonlinear, there are orthogonal basis functions as odd multiples of a power-of-two index form a nonorthogonal subset, but these subsets are orthogonal to one-another.
Decoupling the degrees of freedom consists of showing that $S_{n_0}$ is positive definite, factoring out a large parameter in the action, and integrating by method of steepest descent about $c_{n_0}=0$. We can then neglect terms in the action of lower order in $c_{n_0}$. Requiring $c_{n_0}^2 \ll g$ as $(r,g)\rightarrow(0,0)$, and $c_{n_0}^2 \ll 1/g$ as $(r,g)\rightarrow(0,\infty) \ll 1/g$ allows us to neglect quartic terms containing less than $c_{n_0}^4$ while retaining the quadratic terms as in Eq.(\ref{diagAction}).
At this point it is straightforward to show that $S_{n_0}$ is positive definite as $(r,g)\rightarrow(0,0)$. Let $|r|,g \ll 1/L$ where $L$ is the system size length. Then the positive definite gradient term containing $p_{n_0}^2$ is much greater than the terms containing $r$ and $g$, which guarantees positive definite $S_{n_0}$. When $(r,g)\rightarrow(0,\infty)$ we can use the same argument for $|r|$ and neglect it, but that argument clearly doesn't hold for $g$. Proving positive definiteness of $S_{n_0}$ for any configuration of the degrees of freedom can be a subtle and daunting task, but here we will outline the procedure to lend support to that assumption for our strongly coupled point. In Eq.(\ref{cutoffAction}) the only term that could be negative is 

\beq
\label{crossterm}
2 \int d^d x ~ g \sum_{p\neq q}^{n_{0}} \phi_{n_{0}} \phi_p \phi_q
\eeq
and only when $c_{n_p}$ and $c_{n_q}$ have opposite signs. The factor of 2 accounts for the fact that there are twice as many of each cross-term than each diagonal term in Eq.(\ref{crossterm}). In Eq.(\ref{phiEqn}) negative terms generated by $\phi_{n_0}(-\nabla^2)\phi_{n_0}$ are offset by the larger positive term $g \phi_{n_0}^4$, so we do not need to consider such terms. We find that the largest possible negative contribution to $S_{n_0}$ requires the coefficients of each nonorthogonal subset of $\phi_{n}$ to have alternating signs. WLOG we look at the subset $\phi_1,\phi_3,\phi_5,\ldots$ since the integrals between terms in another subset give identical results. In this case e.g. we have $c_1,c_5,c_9,\ldots>0$ and $c_3,c_7,c_{11},\ldots<0$ and we are interested in calculating cross-terms such as

\begin{widetext}
\beq
0 &>& \int_0^L d^d x ~ c_1 c_3 \text{sn}(4K(m_1)x/L,m_1)\text{sn}(4K(m_3)3x/L,m_3) \\
0 &<& \int_0^L d^d x ~ c_1 c_5 \text{sn}(4K(m_1)x/L,m_1)\text{sn}(4K(m_5)5x/L,m_5).
\eeq
\end{widetext}
As $m_n\rightarrow0$ these terms vanish, but terms such as $\phi_n^2$ approach $1/2$. We find that the relative strength of the negative contributions monotonically increases with increasing $m_n$. Let $g \gg c_n^2 \forall n$ so that $m_n \rightarrow 1 \forall n$. Then all $\text{sn}(4K(m_n)x/L,m_n)$ become square wave functions with wavenumbers $n/L$. The quartic integrals give straightforward analytic results simplified by the fact that the $\phi_{n_0}^2$ contribution is unity and drops out for all integrals and we find

\begin{widetext}
\beq
1 &=&  \lim_{m_n\rightarrow1} \int_0^L d^d x ~ \text{sn}^2(4K(m_n)n x/L,m_n) \\
\frac{\text{GCD}[p,q]}{\text{LCM}[p,q]}  &=& \lim_{m_p,m_q\rightarrow1} \int_0^L d^d x ~ \text{sn}(4K(m_p)p x/L,m_p)\text{sn}(4K(m_q)q x/L,m_q).
\eeq
\end{widetext}
We can calculate the total contribution of cross-terms involving $\phi_p$ as $\sum_q\int\phi_p \phi_q$. Let $p=1$ then this sum becomes

\beq
1-\frac{\pi}{4} &=& \frac{1}{3} - \frac{1}{5} + \frac{1}{7} -\cdots
\eeq
converging to less than unity. Sums for any $p$ are all less than $1/4$, whereas $\int\phi_p^2=1 \forall p$. Recalling that there are twice as many cross-terms than $\phi_n^2$ terms we multiply these sums by a factor of $2$, but they are then all less than $1/2$ ensuring $S_{n_0}$ is positive definite as desired. Our imposed finite cutoff will truncate the series, disallowing arbitrary rearrangement of the original series, and in such a case the largest value of the partial sum is when it reaches the positive unique value of $1/3$. The largest such result occurs locally for $\phi_{n!!}$ and does not exceed $\sqrt{e\pi/2}~\text{Erf}(1/\sqrt{2})-1 \approx 0.4107$. Doubling this value it remains below unity and positive definiteness holds. In order to use the method of steepest descent we must define a large parameter factored out of the action. As $(r,g)\rightarrow(0,0)$ we can factor out the cutoff index $n_0^2$ while requiring that the cutoff wavenumber $n_0/L=1$. We then allow the system size $L$ to increase without limit. As $(r,g)\rightarrow(0,\infty)$ we can simply factor $g$ our of the action as our large value.

\end{document}